\documentclass[11pt]{article}%
\usepackage{amsfonts}
\usepackage{amsmath}
\usepackage{amssymb}
\usepackage{graphicx}%
\newtheorem{theorem}{Theorem}
\newtheorem{acknowledgement}[theorem]{Acknowledgement}

\newtheorem{corollary}[theorem]{Corollary}

\newenvironment{proof}[1][Proof]{\noindent\textbf{#1.} }{\ \rule{0.5em}{0.5em}}
\begin{document}

\title{Quantum States and Hardy's Formulation of the Uncertainty Principle : a
Symplectic Approach}
\author{Maurice de Gosson \ \ Franz Luef\\Faculty of Mathematics, University of Vienna \\Nordbergstrasse 15 \\A-1090 Vienna\\maurice.de.gosson@univie.ac.at\\franz.luef@univie.ac.at}
\maketitle

\begin{abstract}
We express the condition for a phase space Gaussian to be the Wigner
distribution of a mixed quantum state in terms of the symplectic capacity of
the associated Wigner ellipsoid. Our results are motivated by Hardy's
formulation of the uncertainty principle for a function and its Fourier
transform. As a consequence we are able to state a more general form of
Hardy's theorem.

\end{abstract}

\footnotetext[1]{The authors were supported under the EU-project
MEXT-CT-2004-51715.}


\section{Introduction}

In the early days of quantum mechanics Heisenberg made the fundamental
observation that the position of an electron and its momentum cannot be
measured simultaneously with arbitrary precision. Since then many attempts
have been undertaken to turn Heisenberg's uncertainty principle into rigorous
mathematical theorems. The most well-known interpretation is due to Born which
expresses Heisenberg's principle in terms of non-commutativity of a pair of
operators. More precisely, defining the \textit{position operator} $X$ and
\textit{momentum operator} $P$ by $X\psi=x\cdot\psi$ and $P\psi=-i\hbar
\partial_{x}\psi$ ($\psi$ in some adequate dense subspace of $L^{2}%
(\mathbb{R})$) Heisenberg's uncertainty principle is reflected by the
non-commutativity of the position and momentum operator,
\[
XP-PX=i\hbar I.
\]
In his trailblazing work on the mathematical foundations of quantum mechanics
Weyl, inspired by Born's probabilistic interpretation of physical states in
quantum mechanics, showed that the non-commutativity of position and momentum
operator is actually a statement about the variances of $X$ and $P$:
\begin{equation}
\left(  \int_{-\infty}^{\infty}x^{2}|\psi(x)|^{2}dx\right)  ^{1/2}\left(
\int_{-\infty}^{\infty}\hbar^{2}|\partial_{x}\psi(x)|^{2}dx\right)  ^{1/2}%
\geq\tfrac{1}{2}\Vert\psi\Vert^{2}. \label{UP:Weyl}%
\end{equation}
Recently one of us \cite{go05,go06} has pointed out a formulation of the
uncertainty principle in terms of covariance matrices which has several
attractive features and consequences. We will return to this fact later since
it is one of our tools to extend Hardy's theorem to the higher-dimensional
setting. We continue with our short historical overview. Wiener observed that
Weyl's formulation of Heisenberg's uncertainty principle means that a quantum
state $\psi$ and its Fourier transform cannot both be well-localized in phase
space. In \cite{ha32} Hardy obtained the following theorem which turned
Wiener's observation into a rigorous mathematical statement. Defining the
Fourier transform of $\psi\in L^{2}(\mathbb{R})$ by
\[
{\mathcal{F}}\psi(p)=\left(  \tfrac{1}{2\pi\hbar}\right)  ^{1/2}\int_{-\infty
}^{\infty}e^{-\frac{i}{\hbar}px}\psi(x)dx
\]
and noting that $\psi_{0}(x)=e^{-x^{2}}$ is a minimizer of (\ref{UP:Weyl}),
Hardy suggested to measure the localization of $\psi$ and ${\mathcal{F}}\psi$
with respect to a Gaussian:

\begin{theorem}
[Hardy, \cite{ha32}]\label{thm:Hardy} Let $\psi$ be in $L^{2}(\mathbb{R})$. If
there exist constants $C_{X},C_{P}>0$ and $a,b>0$ such that
\begin{equation}
|\psi(x)|\leq C_{X}e^{-\frac{a}{2\hbar}x^{2}}~~\text{and}~~|{\mathcal{F}}%
\psi(p)|\leq C_{P}e^{-\frac{b}{2\hbar}p^{2}} \label{hardy}%
\end{equation}
then: (i) If $ab=1$, there exists $C\in\mathbb{C}$ such that $\psi
(x)=Ce^{-\frac{a}{2\hbar}x^{2}}$. (ii) If $ab>1$, then $\psi$ vanishes
identically. (iii) If $ab<1$ hen the set of functions satisfying (\ref{hardy})
is non-empty (it contains all conveniently rescaled Hermite functions).
\end{theorem}

Hardy's theorem has been generalized and extended to various settings in
mathematics and physics (with, for instance, the Gaussian replaced by some
other exponential functions, and the phase space by some Lie group: see
\cite{th04-1}). Despite the vast literature on Hardy's formulation of
Heisenberg's uncertainty principle we are not aware of any approach which
provides an explanation of the parameters $a$ and $b$. In the present paper we
discuss Hardy's theorem in terms of symplectic geometry. Therefore we are able
to invoke notions of symplectic topology such as the symplectic capacity of a
phase space ellipsoids.

We will do the following in this Letter:

\begin{itemize}
\item We will formulate Hardy's theorem in terms of the notion of symplectic
capacity $c(\mathcal{B}_{M})$ of a phase-space ellipsoid $\mathcal{B}%
_{M}:Mz\cdot z\leq1;$ that capacity is expressed in terms of an invariant
associated to the Williamson diagonal form of $M$;

\item We will apply this result to a characterization of the (cross)-Wigner
distribution of Gaussian states which are localized on an ellipse in phase
space, and obtain an useful estimate. As a by-product we will prove that a
Wigner distribution can never have compact support.
\end{itemize}

\noindent\textbf{Notation}. The symplectic product of two vectors $z=(x,p)$,
$z^{\prime}=(x^{\prime},p^{\prime})$ in $\mathbb{R}^{2n}$ is $\sigma
(z,z^{\prime})=p\cdot x^{\prime}-p^{\prime}\cdot x$ where the dot $\cdot$ is
the usual (Euclidean) scalar product on $\mathbb{R}^{n}$. In matrix notation:%
\[
\sigma(z,z^{\prime})=(z^{\prime})^{T}Jz\text{ \ , \ }J=%
\begin{pmatrix}
0_{n\times n} & I_{n\times n}\\
-I_{n\times n} & 0_{n\times n}%
\end{pmatrix}
\text{.}%
\]
The corresponding symplectic group is denoted by $\operatorname{Sp}%
(n,\mathbb{R})$: the relation $S\in\operatorname{Sp}(n,\mathbb{R})$ means that
$S$ is a real $2n\times2n$ matrix such that $\sigma(Sz,Sz^{\prime}%
)=\sigma(z,z^{\prime})$; equivalently $S^{T}JS=SJS^{T}=J$.

For $(\psi,\phi)\in L^{2}(\mathbb{R}^{n})\times L^{2}(\mathbb{R}^{n})$ the
Wigner--Moyal transform (or: distribution) $W(\psi,\phi)$ is defined by%

\begin{equation}
W(\psi,\phi)(x,p)=\left(  \tfrac{1}{2\pi\hbar}\right)  ^{n}\int_{\mathbb{R}%
^{n}}e^{-\frac{i}{\hbar}p\cdot y}\psi(x+\tfrac{1}{2}y)\overline{\phi
(x-\tfrac{1}{2}y)}dy; \label{wm}%
\end{equation}
the function $W\psi=W(\psi,\psi)$ is the Wigner transform (or: distribution)
of $\psi$.

For $z_{0}=(x_{0},p_{0})$ the Heisenberg--Weyl operator $\widehat{T}(z_{0})$
is defined by%
\[
\widehat{T}(z_{0})\psi(x)=e^{\frac{i}{\hbar}(p_{0}\cdot x-\frac{1}{2}%
p_{0}\cdot x_{0})}\psi(x-x_{0})\text{.}%
\]

\section{Canonical formulation of the uncertainty principle}

In what follows $A$ and $B$ are two (essentially) self-adjoint operators on
$L^{2}(\mathbb{R}^{n})$ with domains $D_{A}$ and $D_{B}.$ For $\psi\in D_{A}$,
$||\psi||_{L^{2}}=1$ we set $\left\langle A\right\rangle _{\psi}=(A\psi,\psi$
$)_{L^{2}}$.

Assume that the (co-)variances
\begin{gather*}
(\Delta A)_{\psi}^{2}=\left\langle A^{2}\right\rangle _{\psi}-\left\langle
A\right\rangle _{\psi}^{2}\text{ \ , \ }(\Delta B)_{\psi}^{2}=\left\langle
B^{2}\right\rangle _{\psi}-\left\langle B\right\rangle _{\psi}^{2}\\
\Delta(A,B)_{\psi}=\tfrac{1}{2}\left\langle AB+BA\right\rangle _{\psi
}-\left\langle A\right\rangle _{\psi}\left\langle B\right\rangle _{\psi}%
\end{gather*}
exist. Then (see for instance Messiah \cite{me91})
\begin{equation}
(\Delta A)_{\psi}^{2}(\Delta B)_{\psi}^{2}\geq\Delta(A,B)_{\psi}^{2}-\tfrac
{1}{4}\left\langle [A,B]\right\rangle _{\psi}^{2}. \label{uncab}%
\end{equation}
The proof of this inequality is based on the trivial identity%
\[
AB=\tfrac{1}{2}(AB+BA)+\tfrac{1}{2}(AB-BA);
\]
notice that since $(AB+BA)^{\ast}=AB+BA$ and $[A,B]^{\ast}=-[A,B]$ we have
$\Delta(A,B)_{\psi}^{2}\geq0$ and $\left\langle [A,B]\right\rangle _{\psi}%
^{2}\leq0$ so that (\ref{uncab}) implies that
\[
(\Delta A)_{\psi}(\Delta B)_{\psi}\geq-\tfrac{1}{4}\left\langle
[A,B]\right\rangle _{\psi}^{2}\geq0.
\]
It is this weak form of the uncertainty principle, obtained by be neglecting
correlations, that is almost exclusively discussed in the mathematical
literature. This is indeed a pity, because one then loses one of the most
interesting and useful features of the uncertainty principle, namely its
canonical invariance.

Specializing to the case where $A$ and $B$ are the operators $X_{j}$ and
$P_{j}$ defined, for $\psi\in\mathcal{S}(\mathbb{R}^{n})$, by $X_{j}\psi
=x_{j}\psi$ and $P_{j}\psi=-i\hbar\partial_{x_{j}}\psi$. Schr\"{o}dinger's
formulation (\ref{uncab}) of Heisenberg's uncertainty principle becomes in
this case%
\begin{align}
(\Delta X_{j})_{\psi}^{2}(\Delta P_{j})_{\psi}^{2}  &  \geq\Delta(X_{j}%
,P_{j})_{\psi}^{2}+\tfrac{1}{4}\hbar^{2}\text{ , }1\leq j\leq n\label{u1}\\
(\Delta X_{j})_{\psi}^{2}(\Delta P_{k})_{\psi}^{2}  &  \geq\Delta(X_{j}%
,P_{k})_{\psi}^{2}\text{ \ for \ }j\neq k; \label{u2}%
\end{align}
neglecting the covariances $\Delta(X_{j},P_{k})_{\psi}$ for all $j,k$ leads to
the \textquotedblleft naive\textquotedblright\ textbook inequalities
\[
(\Delta X_{j})_{\psi}(\Delta P_{j})_{\psi}\geq\tfrac{1}{2}\hbar.
\]

We are now going to rewrite Schr\"{o}dinger's formulation (\ref{u1}),
(\ref{u2}) of the position-momentum uncertainty principle in an equivalent,
but obviously symplectically covariant, way. For this purpose we introduce the
\textquotedblleft covariance matrix\textquotedblright\
\begin{equation}
\Sigma=%
\begin{pmatrix}
\Delta(X,X)_{\psi} & \Delta(X,P)_{\psi}\\
\Delta(P,X)_{\psi} & \Delta(P,P)_{\psi}%
\end{pmatrix}
\end{equation}
where $\Delta(X,X)_{\psi}=((\Delta X_{j}\Delta X_{k})_{\psi})_{1\leq j,k\leq
n}$, and so on.

\begin{theorem}
\label{thon}The inequalities (\ref{u1}), (\ref{u2}) are equivalent to the
following statement: the Hermitian matrix
\begin{equation}
\Sigma+\frac{i\hbar}{2}J\text{ is positive semi-definite.} \label{sipo}%
\end{equation}

\end{theorem}

\begin{proof}
See Narcowich and O'Connell \cite{naoc86}, Narcowich \cite{na90} and Simon
\textit{et al}. \cite{sisumu87,simudu96}.
\end{proof}

The positive semi-definiteness of $\Sigma+\frac{i\hbar}{2}J$ implies that
$\Sigma$ itself is positive; in fact one can show that $\Sigma$ is even
positive-definite (see Narcowich \cite{naoc86}); this allows us to define the
\textquotedblleft Wigner ellipsoid\textquotedblright%
\begin{equation}
\mathcal{W}_{\Sigma}:\frac{1}{2}\Sigma^{-1}z\cdot z\leq1. \label{wig}%
\end{equation}
We will say that $\mathcal{W}_{\Sigma}$ is \textit{quantum mechanically
admissible} when condition (\ref{sipo}) is satisfied. This is a first step
toward a geometrization of the uncertainty principle. Next step is gladly
taken in the forthcoming section.

\section{Uncertainty and Symplectic Capacities}

A fundamental observation is now that condition (\ref{sipo}) can be very
simply stated in terms of an notion familiar from symplectic topology, namely
the \textquotedblleft symplectic capacity\textquotedblright\ of the Wigner
ellipsoid (\ref{wig}). Recall (see for instance Hofer and Zehnder
\cite{hoze94}) that a symplectic capacity on $(\mathbb{R}^{2n},\sigma)$ is the
assignment to every subset $\Omega$ of $\mathbb{R}^{2n}$ of a number
$c(\Omega)\geq0$, or $+\infty$, such that the following conditions hold:

\begin{enumerate}
\item $c(f(\Omega))=c(\Omega)$ for every symplectomorphism $f$ of
$(\mathbb{R}^{2n},\sigma)$

\item $c(\Omega)\leq c(\Omega^{\prime})$ if $\Omega\subset\Omega^{\prime}$

\item $c(\lambda\Omega)=\lambda^{2}c(\Omega)$ for every $\lambda\in\mathbb{R}$

\item $c(Z_{j}(r))=c(B(r))=\pi r^{2}$.
\end{enumerate}

\noindent In Condition 4, $Z_{j}(r)$ and $B(r)$ are, respectively, the
cylinder $x_{j}^{2}+p_{j}^{2}\leq r^{2}$ and the ball $|z|\leq r$. When we
only allow linear or affine symplectomorphisms in Condition 1, we will talk
about \textit{linear symplectic capacities}. The existence of symplectic
capacities is by no means easy to prove; all known constructions are
notoriously difficult (see Hofer and Zehnder \cite{hoze94}, Polterovich
\cite{po01} for explicit examples of symplectic capacities). This difficulty
is after all not so surprising, since the existence of a single symplectic
capacity is equivalent to Gromov's non-squeezing theorem \cite{gr85}, which is
a deep theorem of symplectic topology, whose proof requires sophisticated
techniques (the theory of pseudo-holomorphic curves). Gromov's theorem says
that there is no symplectomorphism $f$ of $(\mathbb{R}^{2n},\sigma)$ such that
$f(B(R))\subset Z_{j}(r)$ if $r<R$ (that such an $f$ exists if $r\geq R$ is
easy to prove). Defining, for $\Omega\subset\mathbb{R}^{2n}$,
\begin{equation}
c_{\text{G}}(\Omega)=\sup_{f\in Symp(n)}\{\pi R^{2}:f(B(R))\subset\Omega\}
\label{cgr}%
\end{equation}
($Symp(n)$ the set of all symplectomorphisms of $(\mathbb{R}^{2n},\sigma
)$)\ it turns out that $c_{\text{G}}$ indeed is a symplectic capacity
($c_{\text{G}}(\Omega)$ is sometimes called \textquotedblleft Gromov's
width\textquotedblright\ or \textquotedblleft symplectic
area\textquotedblright\ of $\Omega$; it can be proven that $c_{\text{G}%
}(\Omega)$ is the usual area when $n=1$).

While there exist infinitely many symplectic capacities on $(\mathbb{R}%
^{2n},\sigma)$ it turns out that all symplectic capacities agree on
phase-space ellipsoids, and moreover agree with the linear symplectic capacity
obtained by restricting $f$ in (\ref{cgr}) to affine symplectic
transformations:%
\[
c_{\text{lin}}(\Omega)=\sup_{S\in I\operatorname{Sp}(n,\mathbb{R})}\{\pi
R^{2}:S(B(R))\subset\Omega\}
\]
where $I\operatorname{Sp}(n,\mathbb{R})$ is the inhomogeneous symplectic
group. Let us precise this result, and relate it to Williamson's theorem
\cite{wi63} (also see Hofer and Zehnder \cite{hoze94} for an alternative
proof). That theorem says that one can diagonalize a positive-definite form
using a symplectic matrix:

\begin{theorem}
[Williamson]Let $M$ be a positive definite $2n\times2n$ real matrix and
$Q(z)=Mz\cdot z$ the associated real quadratic form on $\mathbb{R}^{2n}$. Then
there exists a symplectic matrix $S\in\operatorname*{Sp}(n,\mathbb{R})$ such
that
\begin{equation}
Q(Sz)=\sum_{j=1}^{n}\lambda_{j}(x_{j}^{2}+p_{j}^{2}) \label{q1}%
\end{equation}
the positive numbers $\lambda_{j}$ being the moduli of the eigenvalues $\pm
i\lambda_{j}$ of $JM$.
\end{theorem}

It is customary to order the $\lambda_{j}$ decreasingly: $\lambda_{1}%
\geq\lambda_{1}\geq\cdots\geq\lambda_{n}$ and to call the finite sequence
\[
\operatorname*{Spec}\nolimits_{\sigma}(M)=(\lambda_{1},...,\lambda_{n})
\]
the \textit{symplectic spectrum }of $Q$ (or of $M$). One proves the following
properties:%
\begin{equation}
\operatorname*{Spec}\nolimits_{\sigma}(M^{-1})=(\lambda_{n}^{-1}{}%
...,\lambda_{1}^{-1}) \label{spec1}%
\end{equation}
and%
\begin{equation}
M\leq M^{\prime}\Longrightarrow\operatorname*{Spec}\nolimits_{\sigma}%
(M)\leq\operatorname*{Spec}\nolimits_{\sigma}(M^{\prime}) \label{spec2}%
\end{equation}
(see \textit{e.g.} de Gosson \cite{go05}, Appendix, for a proof). The
diagonalizing symplectic matrix $S$ is of course not unique in general;
however (ibid.) if $S^{\prime}$ is a second diagonalizing matrix then there
exists $U\in\operatorname{Sp}(n,\mathbb{R})\cap O(2n)$ such that $S^{\prime
}=SU$ (or $US$).

It easily follows from Williamson's theorem and the properties of the
symplectic spectrum that the linear symplectic capacity of the ellipsoid
$\mathcal{Q}:Q(z)\leq1$ is $c_{\text{lin}}(\mathcal{Q})=\pi/\lambda_{1}$ and
hence, since all symplectic capacities of an ellipsoid are equal,
\begin{equation}
c(\mathcal{Q})=\frac{\pi}{\lambda_{1}}. \label{cbm}%
\end{equation}

With this terminology the strong uncertainty principle can be restated in the
following very concise form:

\begin{theorem}
\label{th1}The uncertainty principle (\ref{u1})-(\ref{u2}), and hence
condition (\ref{sipo}), are equivalent to the inequality
\begin{equation}
c(\mathcal{W}_{\Sigma})\geq\frac{1}{2}h \label{cw}%
\end{equation}
where $c$ is any symplectic capacity (linear, or not) on $(\mathbb{R}%
^{2n},\sigma)$ and $\mathcal{W}_{\Sigma}$ is the Wigner ellipsoid (\ref{wig}).
\end{theorem}

\begin{proof}
It is a consequence of Williamson's theorem; see de Gosson \cite{go05,go06}.
\end{proof}

The inequality (\ref{cw}) tells us, in particular, that there exists
$S\in\operatorname*{Sp}(n,\mathbb{R})$ sending the ball $B(\sqrt{\hbar}))$ in
$\mathcal{W}_{\Sigma}$. The area of the section of $S(B(\sqrt{\hbar}))$ by any
plane of conjugate variables (or, more generally, by any symplectic plane) is
equal to $\pi\hbar=\frac{1}{2}h$: this is due to the fact that the restriction
of a symplectomorphism to a symplectic subspace is still a symplectomorphism.
As a consequence, we obtain the following geometric formulation of the
uncertainty principle:

\begin{quotation}
\textit{The intersection of a Wigner ellipsoid such that (\ref{cw}) holds by
any symplectic plane cannot be inferior to} $\frac{1}{2}h$.
\end{quotation}

It is interesting to note that we have the following dynamical interpretation
of the results above. The Wigner ellipsoid $\mathcal{W}_{\Sigma}$ can be
viewed as the energy shell $H^{-1}(E)$ for the Hamiltonian function
$H=\frac{1}{2}\Sigma^{-1}z\cdot z$ corresponding to the value $E=1$.
Symplectic capacities being symplectic invariants, it is no restriction to
assume that $\Sigma$ is in Williamson diagonal form, so that
\[
H(z)=\sum_{j=1}^{n}\frac{\omega_{j}}{2}(x_{j}^{2}+p_{j}^{2})
\]
where $(\omega_{1},...,\omega_{n})$ is the symplectic spectrum of $\Sigma
^{-1}$. With these notations we have $c(\mathcal{W}_{\Sigma})=\pi\omega_{1}$
so that the condition $c(\mathcal{W}_{\Sigma})\geq\frac{1}{2}h$ can be
rewritten as%
\[
\oint\nolimits_{\gamma_{1}}pdx\geq\tfrac{1}{2}h
\]
where $\gamma_{1}$ is the shortest periodic orbit carried by $H^{-1}(E)$
(namely, that lying in the $x_{1},p_{1}$ plane).

\section{Hardy's theorem and uncertainty}

In what follows $M$ will always denote a positive-definite real $2n\times2n$
matrix. Recall \cite{li86} that the Wigner transform of a function $\psi\in
L^{2}(\mathbb{R}^{n})$ is defined by%
\[
W\psi(z)=\left(  \tfrac{1}{2\pi\hbar}\right)  ^{n}\int e^{-\frac{i}{\hbar
}p\cdot y}\psi(x+\tfrac{1}{2}y)\overline{\psi(x-\tfrac{1}{2}y)}dy
\]
and that we have%
\begin{equation}
\int W\psi(z)dp=|\psi(x)|^{2}\text{ \ , \ }\int W\psi(z)dx=|{\mathcal{F}}%
\psi(p)|^{2}. \label{marge}%
\end{equation}

The following theorem is a geometric formulation of Hardy's uncertainty principle:

\begin{theorem}
\label{th2}Let $\psi\in L^{2}(\mathbb{R}^{n})$, $\psi\neq0$, and assume that
there exists $C>0$ such that $W\psi(z)\leq Ce^{-\frac{1}{\hbar}Mz\cdot z}$.
Then $c(\mathcal{W}_{\Sigma})\geq\frac{1}{2}h$ where $\mathcal{W}_{\Sigma}$ is
the Wigner ellipsoid corresponding to the choice $\Sigma=\frac{\hbar}{2}%
M^{-1}$ (equivalently $c(\mathcal{B}_{M})\geq\frac{1}{2}h$ where
$\mathcal{B}_{M}:Mz\cdot z\leq\hbar$).
\end{theorem}

\begin{proof}
In view of Williamson's theorem we can find $S\in Sp(n,\mathbb{R})$ be such
that
\[
MSz\cdot Sz=\sum_{j=1}^{n}\lambda_{j}(x_{j}^{2}+p_{j}^{2})
\]
with $\operatorname*{Spec}_{\sigma}(M)=(\lambda_{1},...,\lambda_{n})$ hence
the assumption $W\psi(z)\leq Ce^{-\frac{1}{\hbar}Mz\cdot z}$ can be rewritten
as
\begin{equation}
W\psi(S^{-1}z)\leq C\exp\left(  -\frac{1}{\hbar}\sum_{j=1}^{n}\lambda
_{j}(x_{j}^{2}+p_{j}^{2})\right)  \text{.} \label{maj}%
\end{equation}
Since $W\psi(S^{-1}z)=W\widehat{S}\psi(z)$ where $\widehat{S}$ is any of the
two operators in the metaplectic group $\operatorname*{Mp}(n,\mathbb{R})$ with
projection $S$. Since $\widehat{S}\psi\in L^{2}(\mathbb{R}^{n})$ and
$c(\mathcal{W}_{\Sigma})$ is a symplectic invariant it is no restriction to
assume $S=I$, $\widehat{S}=I$ . Integrating the inequality
\[
W\psi(z)\leq C\exp\left(  -\frac{1}{\hbar}\sum_{j=1}^{n}\lambda_{j}(x_{j}%
^{2}+p_{j}^{2})\right)
\]
in $x$ and $p$, respectively we get, using formulae (\ref{marge}),%
\begin{align}
|\psi(x)|  &  \leq C_{1}\exp\left(  -\frac{1}{2\hbar}\sum_{j=1}^{n}\lambda
_{j}x_{j}^{2}\right) \label{un}\\
|{\mathcal{F}}\psi(p)|  &  \leq C_{1}\exp\left(  -\frac{1}{2\hbar}\sum
_{j=1}^{n}\lambda_{j}p_{j}^{2}\right)  \label{deux}%
\end{align}
for some constant $C_{1}>0$. Let us now introduce the following notation. We
set $\psi_{1}(x_{1})=\psi(x_{1},0,...,0)$ and denote by ${\mathcal{F}}_{1}$
the one-dimensional Fourier transform in the $x_{1}$ variable. Now, we first
note that (\ref{un}) implies that
\begin{equation}
|\psi_{1}(x_{1})|\leq C_{1}\exp\left(  -\frac{\lambda_{1}}{2\hbar}x_{1}%
^{2}\right)  . \label{trois}%
\end{equation}
On the other hand, by definition of the Fourier transform ${\mathcal{F}}$,
\[
\int{\mathcal{F}}\psi(p)dp_{2}\cdot\cdot\cdot dp_{n}=\left(  \tfrac{1}%
{2\pi\hbar}\right)  ^{n/2}\int\int e^{-\frac{i}{\hbar}p\cdot x}\psi
(x)dxdp_{2}\cdot\cdot\cdot dp_{n};
\]
taking into account the Fourier inversion formula this is
\[
\int{\mathcal{F}}\psi(p)dp_{2}\cdot\cdot\cdot dp_{n}=\left(  2\pi\hbar\right)
^{(n-1)/2}{\mathcal{F}}_{1}\psi_{1}(p_{1}).
\]
It follows that%
\[
|{\mathcal{F}}_{1}\psi_{1}(p_{1})|\leq\left(  \tfrac{1}{2\pi\hbar}\right)
^{(n-1)/2}C_{1}\int e^{-\frac{1}{2\hbar}\sum_{j=1}^{n}\lambda_{j}p_{j}^{2}%
}dp_{2}\cdot\cdot\cdot dp_{n}%
\]
that is%
\begin{equation}
|{\mathcal{F}}_{1}\psi_{1}(p_{1})|\leq C_{3}\exp\left(  -\frac{\lambda_{1}%
}{2\hbar}p_{1}^{2}\right)  \label{quatre}%
\end{equation}
for some constant $C_{3}>0$. Applying Hardy's theorem we see that the
condition $\lambda_{1}^{2}\leq1$ is both necessary and sufficient for these
inequalities to hold (remember that $\lambda_{1}\geq\lambda_{2}\geq\cdots
\geq\lambda_{n}$); in view of (\ref{cbm}) this is the same thing as
$c(\mathcal{B}_{M})\geq\frac{1}{2}h$.
\end{proof}

\section{The Case of "Quantum Blobs"}

Let us look at the particular case where the ellipsoid $\mathcal{B}%
_{M}:Mz\cdot z$ is a \textquotedblleft quantum blob\textquotedblright,
\textit{i.e}. the image of the ball $B(\sqrt{\hbar})$ by a linear symplectic
transformation (in which case we have $c(\mathcal{B}_{M})=\frac{1}{2}h$). We
begin by shortly recalling some basic facts about the metaplectic group (for
details and references see for instance Leray \cite{Leray}, or de Gosson
\cite{mdglettmath,Birk}).

The symplectic group $\operatorname{Sp}(n,\mathbb{R})$ is a connected
classical Lie group, contractible to its maximal compact subgroup
$U(n)=\operatorname*{Sp}(n)\cap O(2n,\mathbb{R})$. The latter being
diffeomorphic to $U(n,\mathbb{C})$ we have%
\[
\pi_{1}[\operatorname*{Sp}(n)]\simeq\pi_{1}[U(n,\mathbb{C})]\simeq
(\mathbb{Z},+)
\]
hence $\operatorname*{Sp}(n,\mathbb{R})$ has covering groups
$\operatorname*{Sp}_{q}(n,\mathbb{R})$ of all orders $q=2,3,...,\infty$ and
$\operatorname*{Sp}_{\infty}(n,\mathbb{R})$ is its universal (= simply
connected) covering. A particular role is played by the two-fold covering
$\operatorname*{Sp}_{2}(n,\mathbb{R}),$ because it has a faithful
representation as a subgroup of the unitary group of $L^{2}(\mathbb{R}^{n})$,
the \emph{metaplectic group }$\operatorname*{Mp}(n,\mathbb{R})$. The covering
projection $\pi:\operatorname*{Sp}_{2}(n)\longrightarrow\operatorname*{Sp}(n)$
induces a two-to-one epimorphism $\pi_{\operatorname*{Mp}}:\operatorname*{Mp}%
(n,\mathbb{R})\longrightarrow\operatorname*{Sp}(n,\mathbb{R})$. The two
elements of $\operatorname*{Mp}(n,\mathbb{R})$ that cover $S\in
\operatorname*{Sp}(n,\mathbb{R})$ are denoted by $\widehat{S}$ and
$-\widehat{S}$.

In our context what we will need is the following \textquotedblleft
metaplectic covariance formula\textquotedblright: let $W(\psi,\phi)$\ be the
Wigner--Moyal transform of the pair $(\psi,\phi)\in(L^{2}(\mathbb{R}^{n}%
))^{2}$. Then, for every $S\in\operatorname{Sp}(n,\mathbb{R})$ we have
\begin{equation}
W(\psi,\phi)(S^{-1}z)=W(\widehat{S}\psi,\widehat{S}\phi)(z) \label{metco}%
\end{equation}
where $\widehat{S}\in\operatorname*{Mp}(n,\mathbb{R})$ is such that
$\pi_{\operatorname*{Mp}}(\widehat{S})=S$.

With these notations, Theorem \ref{th2} has the following consequence:

\begin{corollary}
Assume that $\mathcal{B}_{M}=S(B(\sqrt{\hbar}))$ for some $S\in
\operatorname*{Sp}(n,\mathbb{R}).$ If $W\psi(z)\leq Ce^{-\frac{1}{\hbar
}Mz\cdot z}$ then $\psi$ is proportional to the squeezed coherent state
$\widehat{S}^{-1}\psi$ where $\psi_{0}(x)=(\pi\hbar)^{-n/4}e^{-\frac{1}%
{2\hbar}|x|^{2}}$ and $\widehat{S}$ is any of the two metaplectic operators
$\pm\widehat{S}$ covering $S$.
\end{corollary}

\begin{proof}
If $\mathcal{B}_{M}=S(B(\sqrt{\hbar}))$ then $\lambda_{j}=1$ for all
$j=1,...,n$ and the inequality (\ref{maj}) in the proof of Theorem \ref{th2}
can be written $W\psi(S^{-1}z)\leq Ce^{-\frac{1}{\hbar}|z|^{2}}$. Since
$W\psi(S^{-1}z)=W\widehat{S}\psi(z)$ in view of (\ref{metco}), Hardy's theorem
now implies that $\widehat{S}\psi(x)=(\pi\hbar)^{-n/4}e^{-\frac{1}{2\hbar
}|x|^{2}}$ hence our claim.
\end{proof}

\section{Mixed states}

Insofar we have been dealing with pure states; everything actually carries
over without difficulty to the more general case of mixed states. Recall that
a trace-class operator $\widehat{\rho}$ on $L^{2}(\mathbb{R}^{n})$ is called a
\textit{density operator} if it is positive (and hence self-adjoint) and has
trace equal to one. It is advantageous to view $\widehat{\rho}$ as a Weyl
operator, in which case we can write
\[
\widehat{\rho}\psi(x)=\int\int e^{\frac{i}{\hbar}\left\langle
p,x-y\right\rangle }\rho(\tfrac{1}{2}(x+y),p)\psi(y)dydp\text{;}%
\]
the function $\rho$ (which is $(2\pi\hbar)^{n}$ times the symbol of
$\widehat{\rho}$) is called the Wigner distribution of $\widehat{\rho}$. The
average value of a self-adjoint bounded operator $A$ on $L^{2}(\mathbb{R}%
^{n})$ with respect to $\widehat{\rho}$ is then%
\[
\left\langle A\right\rangle _{\widehat{\rho}}=\operatorname*{Tr}(\widehat
{\rho}A)=\int\rho(z)a(z)dz
\]
(see Littlejohn \cite{li86} for a review of the notion). With these notations
Schr\"{o}dinger's form (\ref{uncab}) of the uncertainty principle becomes%
\begin{equation}
(\Delta A)_{\widehat{\rho}}^{2}(\Delta B)_{\widehat{\rho}}^{2}\geq
\Delta(A,B)_{\widehat{\rho}}^{2}-\tfrac{1}{4}\left\langle [A,B]\right\rangle
_{\widehat{\rho}}^{2} \label{uncac}%
\end{equation}
where $(\Delta A)_{\widehat{\rho}}^{2}$, etc. are defined exactly as in the
\textquotedblleft pure\textquotedblright\ case. These definitions extend to
the case where the operators $A$ and $B$ are essentially self-adjoint; in
practice they are required to be defined (at least) on the Schwartz space
$\mathcal{S}(\mathbb{R}^{n})$ of rapidly decreasing functions.

We will assume throughout that the function $z\longmapsto(1+|z|^{2})\rho(z)$
is in $L^{1}(\mathbb{R}^{2n})$; then, in particular, we have the analogues of
the uncertainty inequalities (\ref{u1}), (\ref{u2}):%
\begin{align}
(\Delta X_{j})_{\widehat{\rho}}^{2}(\Delta P_{j})_{\widehat{\rho}}^{2}  &
\geq\Delta(X_{j},P_{j})_{\widehat{\rho}}^{2}+\tfrac{1}{4}\hbar^{2}\text{ ,
}1\leq j\leq n\label{ua}\\
(\Delta X_{j})_{\widehat{\rho}}^{2}(\Delta P_{k})_{\widehat{\rho}}^{2}  &
\geq\Delta(X_{j},P_{k})_{\widehat{\rho}}^{2}\text{ \ for \ }j\neq k.
\label{ub}%
\end{align}

Writing $\Delta(X,X)_{\widehat{\rho}}=((\Delta X_{j}\Delta X_{k}%
)_{\widehat{\rho}})_{1\leq j,k\leq n}$, and so on, we call again the symmetric
$2n\times2n$ matrix%
\begin{equation}
\Sigma=%
\begin{pmatrix}
\Delta(X,X)_{\widehat{\rho}} & \Delta(X,P)_{\widehat{\rho}}\\
\Delta(P,X)_{\widehat{\rho}} & \Delta(P,P)_{\widehat{\rho}}%
\end{pmatrix}
\label{cov}%
\end{equation}
the \textit{covariance matrix} of the density operator $\widehat{\rho}$.

The rub comes from the fact that the positivity of $\rho$ does not guarantee
that $\widehat{\rho}$ is a non-negative operator (this is is a peculiarity of
the Weyl calculus to which much work and effort has been devoted: see for
instance \cite{feph81,fo89} and the references therein). This apparent
difficulty is actually a manifestation of the uncertainty principle; one
proves (Narcowich \cite{na90} and Simon \textit{et al}. \cite{simudu96}) that
a necessary (but not sufficient!) condition for $\widehat{\rho}$\ to be a
density operator is
\begin{equation}
M^{-1}+iJ\text{ is \textit{positive semi-definite.}} \label{semi}%
\end{equation}
A simple calculation shows that we actually have $M=\frac{\hbar}{2}\Sigma
^{-1}$ where $\Sigma$ is the covariance matrix (\ref{cov}) so that condition
(\ref{semi}) is just the strong form (\ref{sipo}) of the uncertainty principle
in Theorem \ref{thon}.

The following result generalizes Theorem \ref{th2} to mixed states:

\begin{theorem}
\label{th3}Let $M>0$ and $\rho$ be a smooth real function on $\mathbb{R}^{2n}$
such that $\int\rho(z)dz=1$. Assume that $\rho(z)\leq Ce^{-\frac{1}{\hbar
}Mz\cdot z}$ for some $C\geq0$ and consider the ellipsoid $\mathcal{B}%
_{M}:Mz\cdot z\leq\hslash$. If $c(\mathcal{B}_{M})<\frac{1}{2}h$ then $\rho$
cannot be the Wigner distribution of any quantum state.
\end{theorem}

\begin{proof}
As in the proof of Theorem \ref{th2} we can assume, taking into account
Williamson's theorem and the invariance of symplectic capacities under
canonical transformations, that
\begin{equation}
\rho(z)\leq\exp\left(  -\frac{1}{\hbar}\sum_{j=1}^{n}\lambda_{j}(x_{j}%
^{2}+p_{j}^{2})\right)  \text{.} \label{rot}%
\end{equation}
Now there exists an orthonormal set of vectors $(\psi_{j})_{j}$ in
$L^{2}(\mathbb{R}^{n})$ and numbers $\alpha_{j}\geq0$, $\sum_{j}\alpha_{j}=1$,
such that
\[
\rho(z)=\sum_{j}\alpha_{j}W\psi_{j}(z)\text{.}%
\]
Integrating the inequality (\ref{rot}) in $x$ and $p$, respectively we thus
have%
\begin{align*}
\sum_{j}\alpha_{j}\int W\psi_{j}(z)dp  &  \leq C_{1}\exp\left(  -\frac
{1}{\hbar}\sum_{j=1}^{n}\lambda_{j}x_{j}^{2}\right) \\
\sum_{j}\alpha_{j}\int W\psi_{j}(z)dp  &  \leq C_{1}\exp\left(  -\frac
{1}{\hbar}\sum_{j=1}^{n}\lambda_{j}p_{j}^{2}\right)  .
\end{align*}
Since $\int W\psi_{j}(z)dp=|\psi_{j}(x)|^{2}$ and $\int W\psi_{j}%
(z)dx=|{\mathcal{F}}\psi_{j}(p)|^{2}$ these inequalities imply in particular
the existence of constants $C_{j}>0$ such that%
\begin{align*}
|\psi_{j}(x)|  &  \leq C_{j}\exp\left(  -\frac{1}{2\hbar}\sum_{j=1}^{n}%
\lambda_{j}x_{j}^{2}\right) \\
|{\mathcal{F}}\psi_{j}(p)|^{2}  &  \leq C_{j}\exp\left(  -\frac{1}{2\hbar}%
\sum_{j=1}^{n}\lambda_{j}p_{j}^{2}\right)  .
\end{align*}
and one concludes as in the proof of Theorem \ref{th2}.
\end{proof}

As a consequence we get the following nonlocality result:

\begin{corollary}
\label{corja}The Wigner distribution $\rho$ of a quantum state cannot have
compact support.
\end{corollary}

\begin{proof}
Suppose that the support of $\rho$ is contained in some ball $B(R)\subset
\mathbb{R}^{2n}$. Let $\lambda$ be a real number such that $0<\lambda$ $<1$.
We can find $C$ $>0$ such that $\rho(z)\leq Ce^{-\frac{1}{\hbar}\lambda
|z|^{2}}$ for all $z$, which contradicts the statement in Theorem \ref{th3}.
\end{proof}

\section{Conclusion and Conjectures}

A question which we have not addressed in this Letter because of lack of space
and timeliness is that of the relation between our constructions and
deformation quantization \cite{BFFLS}. deformation quantization is actually
present everywhere in this work, if only as a watermark. For instance, our
nonlocality statement of Corollary \ref{corja} (a Wigner distribution cannot
have compact support) is related to the nonlocality of the star-product
(intuitively the exponential of a Poisson bracket is a bidifferential operator
of infinite order). This, and many other questions, certainly deserve a
thorough investigation.

On a more pedestrian level, we remark that Gaussians $\psi(x)=Ce^{-\frac
{1}{\hbar}Mz\cdot z}$ for which $\mathcal{B}_{M}$ is a quantum blob correspond
to the ground sate of a generalized harmonic oscillator. A natural question is
whether one could have similar results for the higher modes. The recent work
\cite{bodeja06} of Bonami, or its possible extensions, could play a crucial
role in an answer to these questions.

Another natural extension would be the following: we have been dealing with
non-degenerate Gaussians. It would be interesting to see what happens when $W$
is of the type
\[
W(z)=Ce^{-\frac{1}{\hbar}Mz\cdot z}%
\]
where $M$ is positive semi-definite: $M\geq0$. Williamson's diagonalization
result (\ref{q1}) should then be replaced by the following statement: there
exists $S\in\operatorname*{Sp}(n,\mathbb{R})$ and $k\leq n$, $\ell\leq n-k$
such that%
\[
Q(Sz)=\sum_{j=1}^{k}\lambda_{j}^{\sigma}(x_{j}^{2}+p_{j}^{2})+\sum
_{j=k+1}^{k+\ell}x_{j}^{2}%
\]
where the $\pm i\lambda_{j}^{\sigma}$ ($\lambda_{j}^{\sigma}\geq0$) are the
eigenvalues of $JM$ on the imaginary axis. In this case the inequality
$Q(z)\leq\hbar$ no longer defines an ellipsoid, but rather a phase-space
cylinder; it is easy to calculate the symplectic capacity of this cylinder,
but how can this be related to the question whether $W$ represents the Wigner
distribution of some mixed quantum state?

Finally we wish to mention that Gr\"{o}chenig and Zimmermann have obtained, in
their discussion \cite{grzi01} of uncertainty principles for time-frequency
representations, similar results by completely different methods for the
short-time Fourier transform which is just the matrix coefficient of the
Heisenberg group. Let $f,g\in L^{2}(\mathbb{R}^{n})$. The \textit{short-time
Fourier transform} is defined as follows
\begin{equation}
V_{g}f(x,\xi)=\int_{\mathbb{R}^{n}}e^{-2\pi i\xi\cdot t}f(t)\overline
{g(t-x)}dt \label{stft}%
\end{equation}
They showed that if $f\in S^{\prime}(\mathbb{R}^{n})$, $g\in S^{\prime
}(\mathbb{R}^{n})$ satisfies
\[
|V_{g}f(x,\xi)|=O(e^{-\pi(|x|^{2}+|p|^{2}})
\]
as $|x|,|p|\rightarrow\infty$ then $f$ and $g$ are multiples of $e^{-2\pi
i\xi_{0}t}e^{-\pi(t-x_{0})^{2}}$ for some $(x_{0},\xi_{0})$. The Wigner--Moyal
distribution (\ref{wm}) of $(\psi,\phi)\in S^{\prime}(\mathbb{R}^{n})\times
S^{\prime}(\mathbb{R}^{n})$ and the short-time Fourier transform (\ref{stft})
are related by the formula%
\[
W(\psi,\phi)(x,p)=\left(  \tfrac{2}{\pi\hbar}\right)  ^{n/2}e^{\frac{2\pi
i}{\hbar}p\cdot x}V_{g^{\vee}}f(x\sqrt{2/\pi\hbar},p\sqrt{2/\pi\hbar})
\]
where $f(x)=\psi(x\sqrt{2\pi\hbar})$, $g(x)=\phi(x\sqrt{2\pi\hbar})$, and
$g^{\vee}(x)=g(-x)$, so that Gr\"{o}chenig and Zimmermann's result can be
restated as:

\begin{quotation}
Assume that\ there exists $C>0$ such that
\[
|W(\psi,\phi)(z)|\leq Ce^{-\frac{1}{\hbar}|z|^{2}}\text{ \ for all
}z=(x,p)\text{;}%
\]
then we can find complex constants $C_{\psi}$ and $C_{\phi}$ and $z_{0}%
=(x_{0},p_{0})$ such that $\psi=C_{\psi}\widehat{T}(z_{0})\psi_{0}$ and
$\phi=C_{\phi}\widehat{T}(-z_{0})\psi_{0}$ where $\psi_{0}$ is the standard
coherent state $\psi_{0}(x)=(\pi\hbar)^{-n/4}e^{-\frac{1}{2\hbar}|x|^{2}}$.
\end{quotation}

\begin{acknowledgement}
The authors would like to express their warmest thanks to Professor Daniel
Sternheimer for valuable and interesting comments on earlier versions of this
work, and for having drawn their attention to the relation between this work
and deformation quantization.
\end{acknowledgement}

\end{document}